\documentclass{PoS}

\def\Al{$^{26}$Al}

\def\Ni{$^{56}$Ni}

\def\Co{$^{56}$Co}
\def\Ci{$^{57}$Co}

\def\Fe{$^{56}$Fe}

\def\Fh{$^{60}$Fe}

\def\Ti{$^{44}$Ti}

\def\Ca{$^{44}$Ca}
\def\Fhal{$^{60}$Fe/$^{26}$Al}
\def\degree{\hbox{$^o$}}

\def\ra{$\rightarrow$}
\def\aa{$\alpha$}
\def\ga{$\gamma$}

\def\cs{cm$^{-2}$ s$^{-1}$}
\def\ms{M$_{\odot}$}

\def\ps{e$^+$ s$^{-1}$}

\title{Nucleosynthesis and gamma-ray lines  }

\ShortTitle{Nucleosynthesis and gamma-ray lines  }

\author{\speaker{Nikos Prantzos}\\
        UMR7095 UPMC and Institut d'Astrophysique de Paris\\
        E-mail: \email{prantzos@iap.fr}}

\abstract{Astrophysical gamma-ray spectroscopy is an invaluable tool
for studying nuclear astrophysics, supernova structure,  recent
star formation  in the 
Milky Way and mixing of nucleosynthesis products in the interstellar medium. 
After a short, historical, introduction to the field, 
I present a brief review of the most important current issues. 
Emphasis is given to radioactivities produced by massive stars
and associated supernova explosions, and in particular, those
related to observations  carried out by INTEGRAL:
short-lived \Ti \ from CasA and long-lived
\Al \ and \Fh \ from massive stars. The observed 
 511 keV emission from positron annihilation in the Galaxy 
 and the role of stellar radioactivity and other potential positron
 sources are also discussed.}

\FullConference{8th INTEGRAL Workshop The Restless Gamma-ray Universe- Integral2010,\\
		September 27-30, 2010\\
		Dublin Ireland}

\begin{document}

\section{Historical background}

Gamma-ray line astronomy with cosmic radioactivities 
was essentially founded with
the landmark paper of Clayton, Colgate and Fishman (1969). That work
clarified the implications of the production of \Ni \ (a doubly magic, and 
yet unstable nucleus) during explosive Si-burning in supernovae (SN). 
In particular, it opened
exciting perspectives for $\gamma$-ray line astronomy, by suggesting that
any supernova within the local group of galaxies woud be detectable
in the characteristic $\gamma$-ray lines 
resulting from the radioactive decay of \Ni \ (lifetime $\tau_{Ni-56}$=8.8 d) 
and its daughter nucleus \Co \ ($\tau_{Co-56}$=0.31 y).

In the 70's D. Clayton identified most of the radionuclides of 
astrophysical interest (i.e. giving a detectable $\gamma$-ray line signal);
for that purpose, he evaluated their average SN yields by assuming that the
corresponding daughter stable nuclei are produced in their solar system 
abundances{\footnote {For a vivid account of the history and foundations 
of $\gamma$-ray line astronomy  (and astronomy with radioactivities in general) 
see Chapter 2, written by D. D. Clayton, in the recent monograph  edited by
Diehl, Hartmann and Prantzos (2010).}}. Amazingly enough (or naturally enough, depending on one's point
of view) his predictions of average SN radionuclide yields (Table 2 in
Clayton 1982) are in excellent agreement with modern yield calculations, based
on full stellar models and detailed nuclear physics (see Fig. 1 in Prantzos 2004a). 
Only the
importance of \Al \ ($\tau_{Al-26}$=1.04 10$^6$ y)
escaped Clayton's (1982) attention, perhaps because its
daughter nucleus $^{26}$Mg is produced in its stable form, making the
evaluation of the parent's yield quite uncertain. That uncertainty did not
prevent Arnett (1977) and Ramaty and Lingenfelter (1977) 
from arguing  that, even if only
10$^{-3}$ of solar $^{26}$Mg is produced as \Al, the resulting Galactic
flux from tens of thousands of supernovae (during the $\sim$1 Myr lifetime
of \Al) would be of the order of 10$^{-4}$ \cs.

In the case of \Al \  nature appeared quite generous, providing a \ga-ray
flux even larger than the optimistic estimates of Ramaty and Lingenfelter 
(1977): the HEAO-3 satellite detected the corresponding 1.8 MeV line from 
the Galactic  center direction at a level
of 4 10$^{-4}$ \cs (Mahoney et al. 1984). That detection, the first ever of
a cosmic radioactivity in $\gamma$-rays, showed that nucleosynthesis is still active in the 
Milky Way; however, the implied large amount of galactic \Al \ ($\sim$2 \ms \
per Myr, assuming steady state) was difficult to accomodate in conventional
models of galactic chemical evolution if SN were the main \Al \ source
(Clayton 1984), since $^{27}$Al would be overproduced in that case; however, 
if the  ``closed box model'' assumption is dropped and {\it infall}
is assumed in the chemical evolution model, that difficulty
is removed, as subsequently shown by Clayton and Leising (1987).

Another welcome mini-surprise came a few years later, when the \Co \
\ga-ray lines were detected in the supernova SN1987A, a $\sim$20 \ms \
star that exploded in the Large Magellanic Cloud. On theoretical  grounds,
it was expected that a SNIa (exploding white dwarf of $\sim$1.4 \ms \ that
produces $\sim$0.7 \ms \ of \Ni) would be the first to be detected in \ga-ray 
lines; indeed, the large envelope mass of massive exploding 
stars  ($\sim$10 \ms) allows only small  amounts
of \ga-rays to leak out of SNII, 
making  the detectability of such objects problematic.
Despite the intrinsically weak \ga-ray line emissivity of SN1987A, the 
proximity
of LMC allowed the first detection of the tell-tale \ga-ray line signature from
the  radioactive chain  \Ni\ra\Co\ra\Fe \ (Matz et al. 1988); this confirmed a 20-year 
old conjecture, namely that the abundant \Fe \ is produced in the
form of radioactive \Ni.

Those discoveries laid the observational foundations of the field of \ga-ray
line astronomy with radioactivities. The next steps were made in the 90ies, 
thanks
to the contributions of the Compton Gamma-Ray Observatory (CGRO). First, the
{\it OSSE} instrument aboard CGRO detected the \ga-ray lines of \Ci \ ($\tau_{Co-57}$=1.1 y)  
from SN1987A
(Kurfess et al. 1992); the determination of the
abundance ratio of the isotopes with mass numbers
56 and 57 offered a unique probe of the physical conditions in the innermost
layers of the supernova, where those isotopes are synthesized 
(Clayton et al. 1992). On the other hand, the {\it {\it COMPTEL}} \ instrument 
mapped the
Miky Way in the light of the 1.8 MeV line and found irregular emission
along the plane of the Milky Way and prominent ``hot-spots'' in directions
approximately tangent to the spiral arms (Diehl et al. 1995), which suggests
that massive stars (SNII and/or WR) are at the origin of galactic \Al \ (as
pointed out in  Prantzos 1991, 1993) and not an old stellar population like
e.g. novae or low mass AGB stars. 

Furthermore, {\it {\it COMPTEL}} \  detected the 1.16 MeV line of
 radioactive
\Ti \ ($\tau_{Ti-44}$=89 y)in the Cas-A supernova remnant (Iyudin et al 1994). That discovery
 offered
another valuable estimate of the yield of a radioactive isotope produced
in a massive star explosion (although, in that case the progenitor star mass
is not known, contrary to the case of SN1987A). On the other hand, it also 
created some new problems, since current  models of core collapse supernova
do not seem able to account for the yield inferred from the observations (see Sec. 2 and Fig. 1).

After CGRO and before INTEGRAL, another important discovery was
made in the field: the RHESSI experiment detected the characteristic decay lines
of \Fh \ (Smith 2004), another long-lived isotope ($\tau_{Fe-60}$=3.8 10$^6$ y). 
The  \Fh \ lines were also detected by {\it SPI}/INTEGRAL 
after 5 years of observations, and the observed \Al/\Fh \ flux ratio appears
compatible with theoretical expectations, which are however subject to large
uncertainties yet (see Sec. 3).

Finally, in the past few years, the study of the 511 keV emission  from positron annihilation
in the galaxy attracted particular attention from astronomers and particle physicists. It is the
oldest (Johnston et al. 1972) and brightest 
\ga - ray line detected from outside the solar system, but despite more than 
30 years of study, the origin of the annihilating positrons remains unknown yet (Sec. 4; see also the
recent extensive review of Prantzos et al. 2010).

In the following I shall focus on the 
radioactivities produced by massive stars and associated supernova explosions, 
and in particular, those related to observations  carried 
out by INTEGRAL. Recent reviews of similar scope are provided in  Leising and Diehl (2009)
and Diehl (2009),
while a monograph on "Astronomy with Radioactivities", covering all topics related
to $\gamma$-ray line astronomy, appeared recently (Diehl, Hartmann and Prantzos 2010).

\section{\Ni \ and \Ti \ from core collapse supernovae (CCSN)}

Both \Ni \ and \Ti \ are produced in the innermost layers of core
collapse SN, through explosive Si-burning. 
Their yields (and those of other Fe-peak nuclei) are extremely difficult to 
evaluate from first principles, at least in the framework of current models
of CCSN. The layers undergoing explosive Si-burning are 
very close to the ``mass-cut'', that fiducial surface separating the supernova
ejecta from the material that falls back to the compact object (after the
passage of the reverse shock). Since no consistent model of a core 
collapse supernova explosion exists up to now (e.g. Magkotsios et al. 2010 and refences therein), 
the position of  the mass-cut
is not well constrained.  

The presence of \Ni \ in SN1987A has been unambiguously
inferred from the detection of 847 keV and 1238 keV 
 $\gamma$-ray lines of the decay of its daughter nucleus \Co.
 Their early appearance ($\sim$6 months earlier than expected from 
 spherically symmetric stratified models, from e.g. Gehrels et al. 1987) 
 suggested that the SN ejecta were
 asymmetric, with  \Co \ being driven close to the surface by
 hydrodynamic instabilities. 
The yield of \Ni \ has been estimated from the extrapolation of the early
optical lightcurve to the origin of the explosion (precisely known thanks
to the neutrino signal, see Arnett et al. 1989 and references therein);
the derived value, 0.07 \ms, is often taken as a ``canonical'' one for
CCSN, e.g. in studies of galactic chemical evolution.
It turns out, however, that CCSN display a wide range of \Ni \ 
values, spanning a range  of at least one order of magnitude; it also appears  that there is a clear 
correlation between the amount of \Ni \ and the energy of the explosion (Hamuy 2003)
probably because a shock of larger energy heats a larger amount of
material to NSE conditions.

SN1987A was a "once in a lifetime"event and it is improbable that another CCSN will be seen in the light of the \Co \ lines in the next decades. Prospects are better for thermonuclear supernovae (SNIa),
although none has been seen up to now (see discussion in Leising and Diehl 2009, and references therein). 
Such a detection, combined with an optical one, would allow an unambiguous identification
of the \Ni \ yield. Probing the physics of the explosion will require observations of the \ga - ray
lightcurve, in particular during the period when the SN envelope becomes optically thin (see 
Horiuchi and Beacom 2010 for an updated discussion of the perspectives for such detections).  

\Ti \ has not been directly detected in SN1987A up to now. Modelling of the
late lightcurve of that supernova and of  the infrared emission lines of the ejecta suggests that it may be powered by 
1-2 10$^{-4}$ \ms \ of \Ti \ (Fransson and Kozma 2002, Motizuki and Kumagai 2004).
The expected flux in the high energy 1157 keV line is $\sim$5 10$^{-6}$ ph/cm$^2$/s, i.e. considerably
lower than the $\sim$2 10$^{-5}$ ph/cm$^2$/s sensitivity of {\it SPI} for an
exposure of 1 Ms; it will undoubtedly constitute a major target for the  next $\gamma$-ray satellite
in the MeV range.

The $\gamma$-ray lines of \Ti \ have been detected in the $\sim$320 yr old
CasA supernova remnant, lying 
at a distance of $\sim$3.4 kpc from Earth. Both the
high energy line at 1.157 MeV and the low energy ones, at 68 and 78 keV,
have been detected: the former by {\it COMPTEL} (Iyudin et al. 1994) and
the latter by Beppo-SAX (Vink et al. 2001) and by {\it IBIS}/INTEGRAL (Renaud et al. 2006). 
On the contrary, the 1.157 MeV line was not detected by {\it SPI}/INTEGRAL; taking into account the
aforementioned detections and the energy resolution of {\it SPI}, the non-detection by {\it SPI} constrains the
velocity dispersion of the Ti-rich ejecta to $>$500 km/s (Martin et al. 2009).
The detected flux of 3.3$\pm$0.6 10$^{-5}$ ph/cm$^2$/s
from {\it COMPTEL},  points to a \Ti \ yield
of $\sim$1.7 10$^{-4}$ \ms. Similar values, i.e. 1-2 10$^{-4}$ \ms, are 
obtained through a study of the combined fluxes of the low energy lines
(Vink et al. 2002, Renaud et al. 2006), although the modelisation of the underlying continuum
spectrum makes the analysis very difficult. Note that the CasA
yield of \Ti \ suffers from uncertainties related to the ionisation stage of the
CasA remnant. \Ti \ decays by orbital electron capture and an ionised
medium could slow down its decay (Mochizuki et al. 1999), reducing considerably the derived yield  
(see also Motizuki and Kumagai 2004).

\begin{figure}[t!]
\centering
\includegraphics[height=0.73\textwidth,angle=-90]{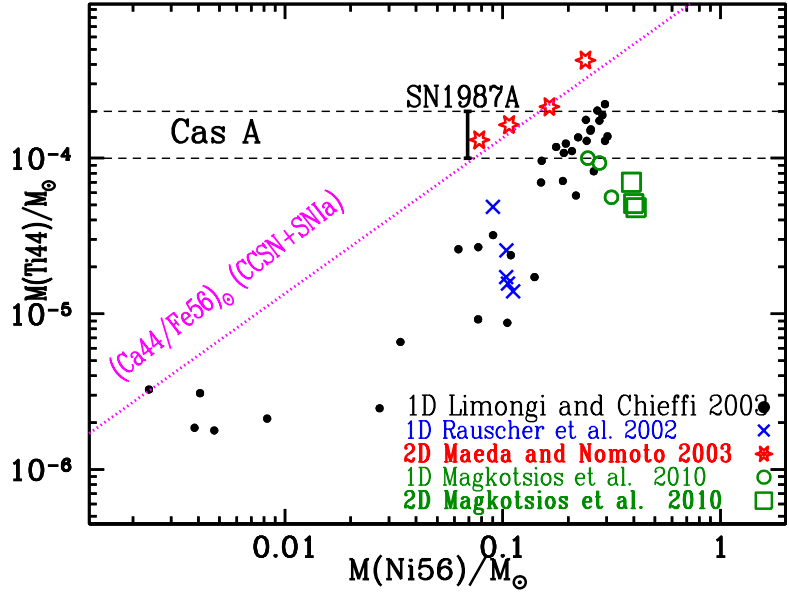}
\caption{Yield of \Ti \ vs yield of \Ni, from models and observations.
Model results are from Limongi and Chieffi (2003, filled circles, with 
large variations in yields due to variations in both stellar mass - from
15 to 35 \ms \ - and
explosion energy), Rauscher et al. (2002, crosses, for stars in the
15 to 25 \ms \ range and explosion energies of 10$^{51}$ ergs),
Maeda and Nomoto (2003, asterisks, for axisymmetric explosions in
25 and 40 \ms \ stars, producing high \Ti/\Ni \ ratios) and Magkotsios et al. (2010,
open circles for 1D and open squares for 2D models). Estimated amount of
\Ti \ detected in CasA appears between 
horizontal dashed lines (assuming that its decay rate has not been
affected by ionisation in the CasA remnant).
The amount of \Ti \ in SN1987A is deduced from its late optical lightcurve.
The diagonal dotted line indicates the solar ratio of the corresponding
stable isotopes ($^{44}$Ca/ $^{56}$Fe)$_{\odot}$.  
\label{fig:TivsNi}}
\end{figure}



In summary: from optical observations we have a wide range of values for the
\Ni \ yields of core collapse SN, and a precise value of 0.07 \ms \ 
for SN1987A; and for \Ti \ yields we have similar values, i.e. 1-2 10$^{-4}$
\ms,  for both SN1987A (indirectly, through the modelisation of the UVOIR 
light)  and for CasA (directly, through $\gamma$-ray lines,
albeit with a systematic uncertainty resulting from poorly constrained
ionisation effects). How do these observations compare to theory ?

The results of recent  calculations are plotted as \Ti \ yield vs \Ni \ yield
in Fig. 1, where the solar
ratio of the corresponding stable isotopes is also displayed as a diagonal 
line. With one exception (to be discussed below) none of the theoretical
results matches the SN1987A value of the \Ti/\Ni \ ratio.
In fact, those results are  $\sim$3 times lower than the solar ratio of 
(\Ca/\Fe)$_{\odot} \sim$10$^{-3}$. This 
implies that such explosions cannot produce the solar \Ca, since
\Fe \ would be overproduced in that case (e.g. Timmes et al.
1996). Moreover, there is another important source of Fe,
SNIa, which  produce about 0.5-0.65 of solar
\Fe, but very little \Ca; this makes the defficiency of \Ca \ from 
CCSN even more serious than appearing in Fig. 1, since it implies
that CCSN {\it should} 
produce a \Ti/\Ni \ ratio {\it at least  twice} 
solar in order to compensate for the \Fe \ production of SNIa (Prantzos 2004a).

In the case of CasA, the \Ni \ yield is not known, but it is constrained from
the non-detection of the CasA explosion in the 1680's (Hartmann et al. 1997), which
suggests a  \Ti/\Ni \ ratio at least as high as in SN1987A, substantially
highrer than obtained in most models. It has been argued that such high ratio
may be obtained in multi-dimensional, aspherical,  models of energetic explosions
of rotating massive stars (hypernovae): material along the
jet (rotation) axis undergoes higher temperatures and entropies (i.e. lower densities)
than material in normal spherical explosions, resulting in the production
of large \Ti \ amounts and  \Ti/\Ni \ ratios (Maeda and Nomoto 2003). 
However, recent hydrodynamic simulations for rotating CCSN do not confirm
that finding (Magkotsios et al. 2010 and Fig. 1). Thus, although there
is observational evidence for asphericity in both SN1987A (Wang et al. 2002) and
CasA (Schure et al. 2008 and references therein), it is not clear whether
this property helps with the \Ti/\Ni \ ratio.

The difficulty of present day CCSN nucleosynthesis models to produce
sufficiently high \Ti/\Ni \ ratios also bears to another issue: searches
of the Milky way with HEAO-3, SMM, COMPTEL and INTEGRAL failed to detect
other \Ti \ sources than CasA up to now, although a few sources are
expected on the basis of inferred \Ti \ yields and Galactic CCSN frequency
(The et al. 2006, Renaud et al. 2006). This may suggest that the main source
of $^{44}$Ca in the Galaxy may be a rare type of SN, of high \Ti \ yield
($\sim$10$^{-3}$ \ms), e.g. He-triggered SNIa of low mass (Woosley et al. 1986).

\section{\Al \ and \Fh \ from massive stars}

\Al \ is the first radioactive nucleus ever detected in the Galaxy
through its characteristic gamma-ray line signature, at 1.8 MeV
(Mahoney et al. 1984). Since its lifetime of $\sim$1
Myr is short w.r.t. galactic evolution timescales, 
its detection convincingly demonstrates that nucleosynthesis
is still active in the Milky Way (Clayton 1984).

\begin{figure}
\includegraphics[width=0.49\textwidth]{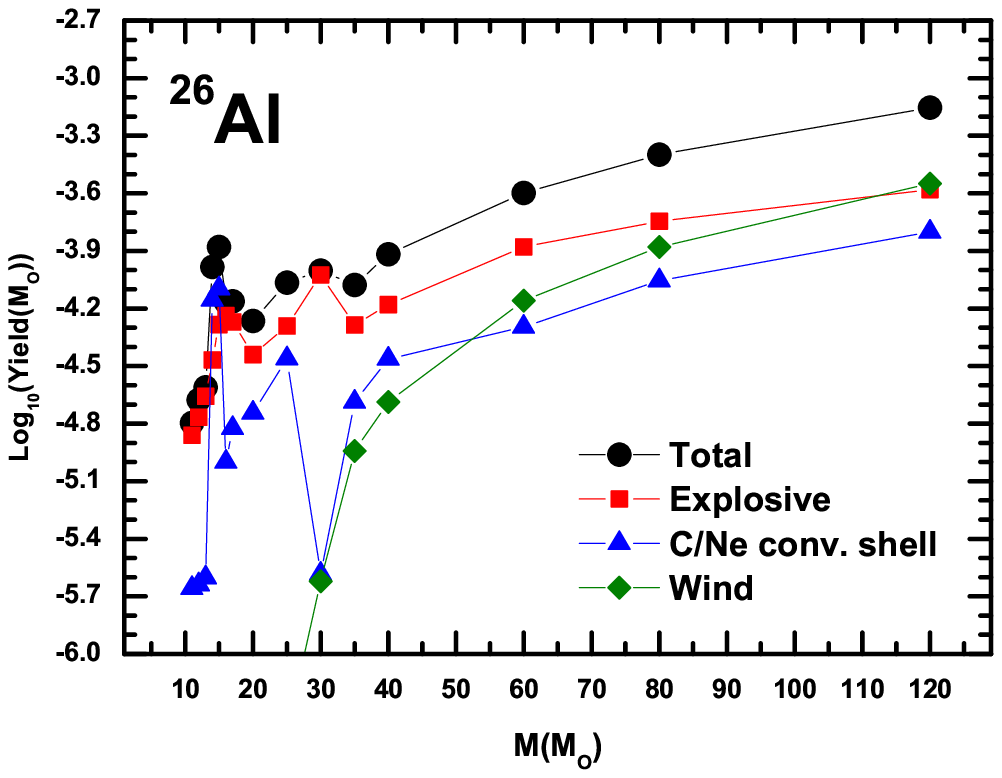}
\qquad   
\includegraphics[width=0.49\textwidth]{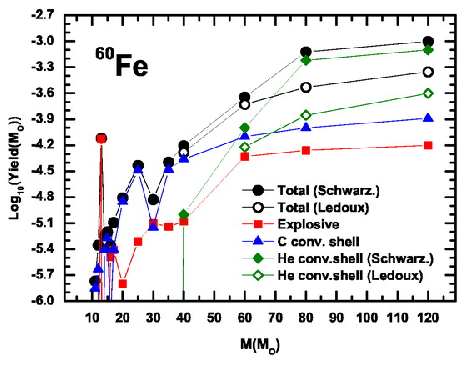}
\caption{Yields of $^{26}$Al (right) and $^{60}$Fe (left) from massive, mass losing  stars of
solar metallicity, according to
Chieffi and Limongi (2006). \Al \ yields  are dominated by explosive nucleosynthesis, while those
of $^{60}$Fe by hydrostatic production in the He-shell. }
\label{fig:AlFe_yields}
\end{figure}

The morphology of the 1.8 MeV emission, as established by 
{\it COMPTEL}/CGRO and {\it SPI}/INTEGRAL
clearly suggests a young population at the origin of \Al, since it is concentrated along
the plane of the Galactic disk. The degree of the irregularity ("patchiness") of that
emission is not well established yet, since it depends on the method of analysis
(i.e. Plushke et al. 2001 vs Kn\" odlseder 1999). Although it is tempting to 
identify some of the "hot-spots" seen in the COMPTEL map with tangents to the
spiral arms (as predicted in Prantzos 1991, 1993) only the star forming regions of
Cygnus  (Kn\" odlseder 2000) and Sco-Cen (Diehl et al. 2010) are unambiguously identified up to now.

For several years, progress has been hampered by the difficulty to evaluate distances
to the regions of the 1.8 MeV emission, which could be dominated by nearby sources. The
high resolution Ge spectrometer of {\it SPI} allowed for the first time to measure Doppler
shifts and derive radial velocities of the emitting regions (a technique widely
used in radioastronomy to map 21 cm emission of HI): the results are consistent with expectations from
large scale rotation of the galactic disk (Kretschmer et al. 2010) and implies that
most of \Al \ is moving as the average ISM. This allows, in turn, to use geometrical
models of the large scale distribution of the ISM (normalised to the mesured 1.8 MeV flux),
to derive the total mass of \Al, which is 2.7$\pm$0,7 \ms \ (Wang et al. 2009).
Moreover, the observed broadening of the 1.8 MeV line is consistent with expectations from
Galactic rotation and suggests that \Al \ is at rest with respect to the ISM (at least
in the plane of the disk). 

The most plausible sources for the inferred $\sim$2 \ms/Myr of \Al \ 
(assuming a steady state between its
production and radioactive decay in the ISM) are massive stars{\footnote {Massive AGB stars (5-8 \ms) 
cannot be excluded, but their \Al \ yields are difficult to evaluate and appear small.}}. The
roles of their winds (expelling \Al \ from hydrostatic H-burning) and explosions (expelling \Al \
from subsequent nuclear burning phases) remained unclear for two decades. Chieffi and Limongi (2006), using non-rotating models of mass losing stars of solar metallicity, 
found that explosive yields always dominate (Fig. 2 left). One should keep in mind, however,
that substantial uncertainties (related to convective mixing or nuclear reaction rates, e.g. Tur et al. 2010)
still affect the \Al \ yields, while rotation and higher metallicities (as appropriate for the
inner Galaxy) might affect the relative importance of hydrostatic vs explosive yields.

The original aims of $\gamma$-ray line astronomy, as formulated in e.g. Clayton (1982)
concerned the study of nucleosynthesis and SN structure, through observations
requiring high energy resolution. The spatial resolution
of satellite instruments made it possible to tackle new issues, related to large scale star formation
(in stellar associations or the whole Galaxy)
and mixing of nucleosynthesis products in the ISM. Thus, Diehl et al. (2006) used the total Galactic 
\Al \ flux, combined to theoretical \Al \ yields, to infer a rate of 1.9$\pm$1.1 CCSN in the Galaxy
(consistent with more conventional estimates), while  Martin et al. (2010) and Voss et al. (2010)
 studied recently \Al \ production and evolution in Cygnus and Orion, respectively, with population synthesis models. On the other hand, preliminary constraints on
the vertical extent of the \Al \ distribution in the Galaxy (and, thereof, on the existence
of galactic "fountains"or "chimneys") can  be obtained from the study of the latitude extent 
of the 1.8 MeV emission (Wang et al. 2009, see Fig. 3, right).

\begin{figure}
\includegraphics[width=0.51\textwidth]{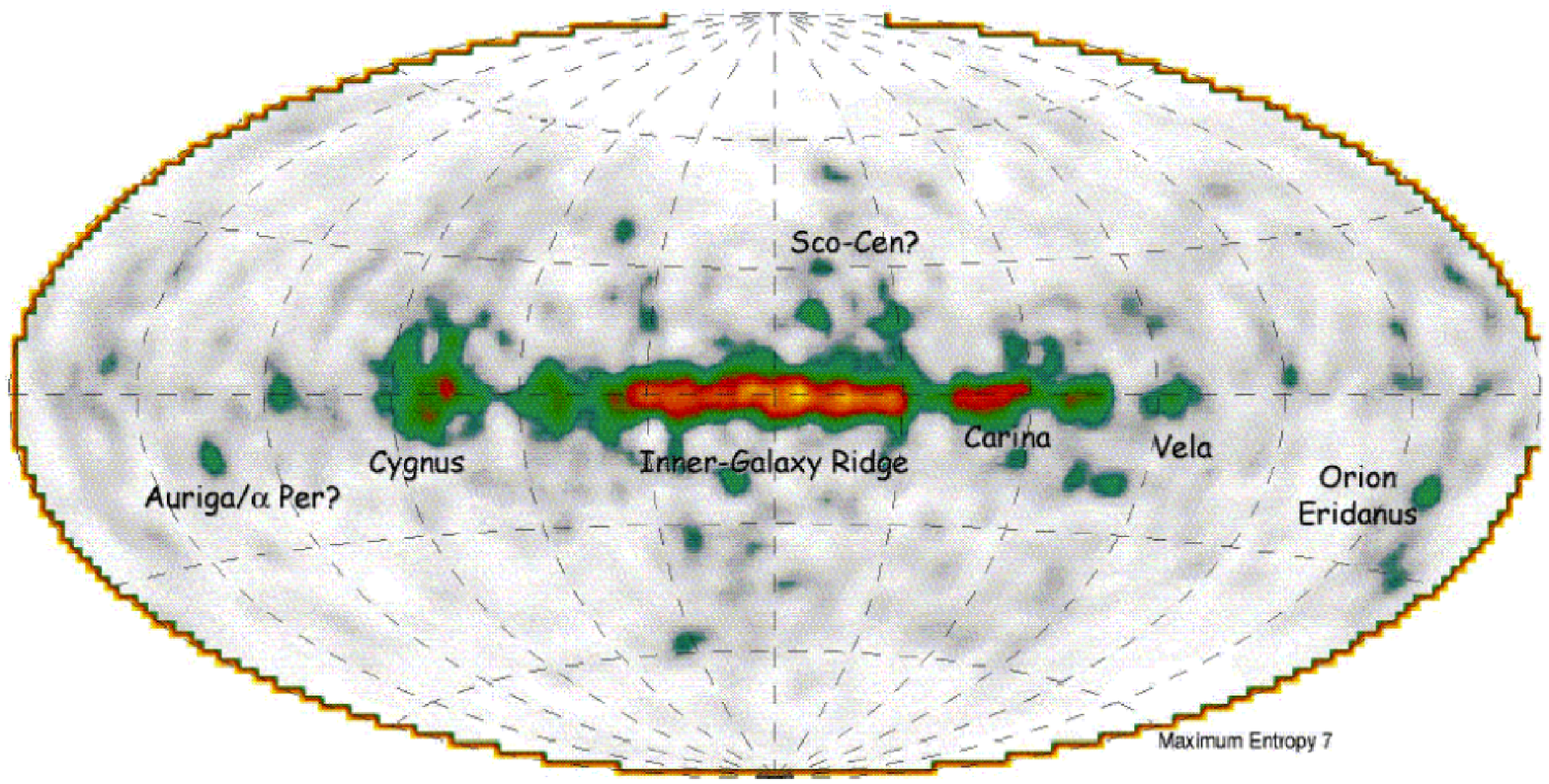}
\qquad
\includegraphics[width=0.46\textwidth]{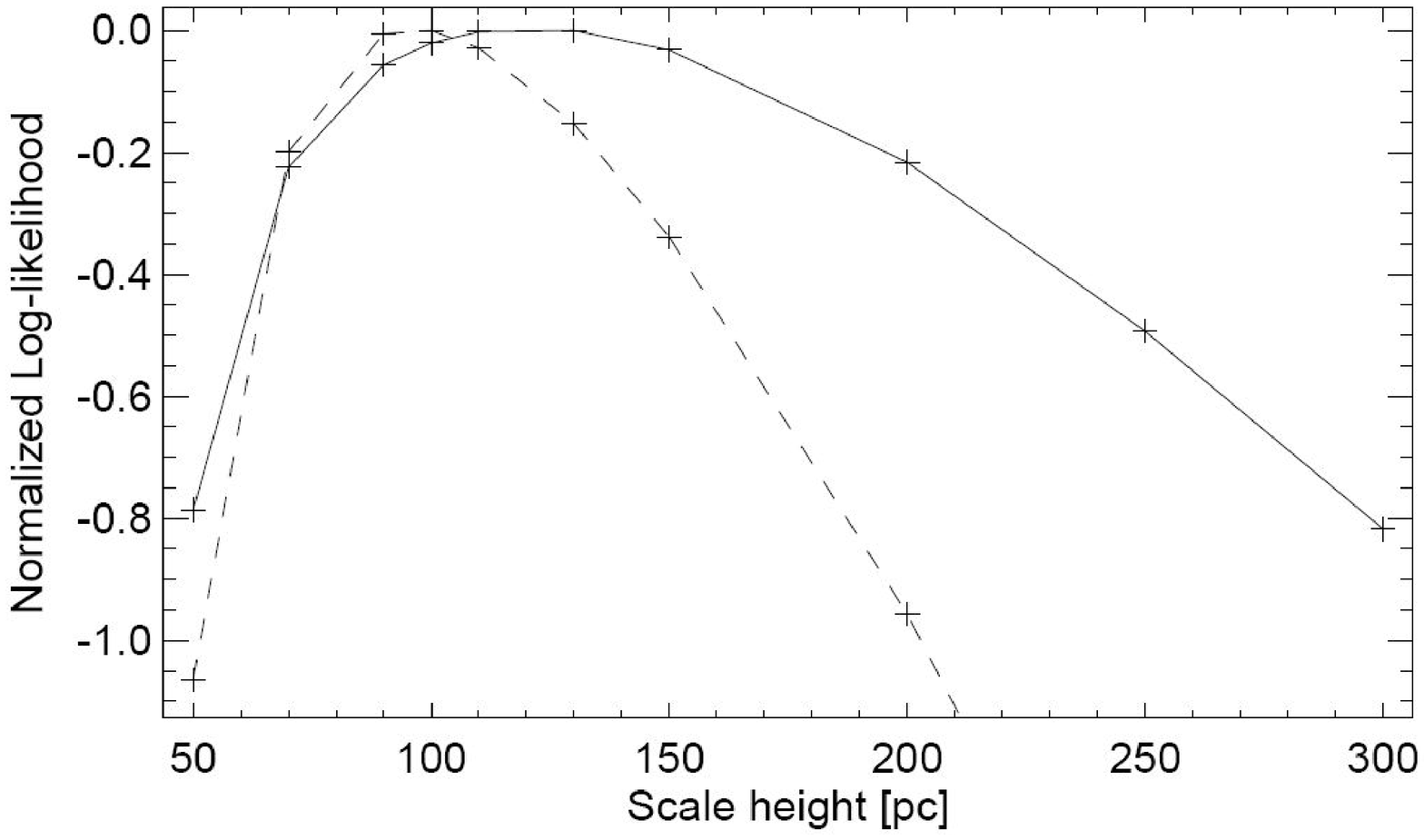}
\caption{Right: COMPTEL map of Galactic \Al (from Pluschke et al. (2001). Left: Evaluation
of the scaleheight of the \Al \ distribution of the Galaxy, from the estimated latitude extent of 
the 1.8 meV emission (from Wang et al. 2009). }
\label{fig:Almap}
\end{figure}

Clayton (1982)  pointed out that SNII explosions produce \Fh,
a radioactivity, with a lifetime comparable to the one of \Al{\footnote{The most recent
measurements of \Fh \ lifetime give $\tau_{Fe-60}$=3.78$\pm$0.06 Myr (Rugel et al. 2009),
a value almost twice as large as previously thought.}. The 
gamma-ray line flux ratio of \Fhal \ in the Galaxy (assuming both radioactivities
in steady state) would provide then a "clean" probe
of stellar nucleosnthesis, unbiased by e.g. absolute values of CCSN rates.
Based on  calculations
from Woosley and Weaver (1995, with models having no mass loss or rotation) 
Timmes et al. (1995) found that the expected ratio of \Fhal \ from CCSN
(for each of the  two lines of \Fh) is 0.16.

The \Fh \ lines were detected by RHESSI (Smith 2004) and subsequently confirmed by
{\it SPI}/INTEGRAL (Harris et al. 2005; Wang et al. 2007, see Fig. 4 left). 
The reported {\it SPI} flux ratio is
\Fhal=0.14$\pm$0.06, but potentially important systematic effects 
(from nearby instrumental lines) cannot be axcluded.
Taken at face value, the reported ratio is in astonishingly good
 agreement with original
predictions. However, in the meantime, refined 
theoretical models predicted substantially
 higher \Fhal \ values, (more \Fh \ and less \Al) 
 as pointed out in Prantzos (2004b). The most recent
 works in the field (Woosley and Heger 2007, Chieffi and Limongi 2006) still 
 predict values on the high side of the {\it SPI} result, at least for plausible values
 of various physics inputs (e.g. Fig. 4 right). 
 
 It is clear, however, 
 that substantial uncertainties still
 remain in stellar and nuclear physics, both for \Al \ (see above)
 and \Fh. The latter is produced mostly
 by hydrostatic burning in the He- and C- layers{\footnote{Notice that
 a large fraction of \Fh \ in the bottom of the He-layer is produced {\it after}
 central O-burning, which implies that virtualy all stages of stellar
 evolution are important for \Fh \ production (Tur et al. (2010).}}, through neutron
 captures on Fe-seed nuclei (Fig. 2 right). 
 Convection (still a major unknown in stellar
 evolution calculations) plays a key role in determining the sizes
 of those convective shells, but other factors, like mass loss (for the
 most massive stars, e.g. above 50 \ms) or rotation also turn out to 
be important. Besides, uncertainties on nuclear reaction rates, including
reactions which are not directly involved in production/destruction
of \Fh, like e.g. the 3-alpha or $^{12}$C($\alpha,\gamma$)$^{16}$O rates,
may greatly affect the final yield of \Fh \ (see Tur et al. 2010).
This leaves a lot of theoretical issues unsettled yet and underscores the
importance of $\gamma$-ray line observations in the Galaxy, both at large
scale and at smaller scales (to determine any gradient of the \Fhal \ ratio,
or its value in young star forming regions like Cygnus, still unaffected
from the action of CCSN).

\begin{figure}
\includegraphics[width=0.48\textwidth]{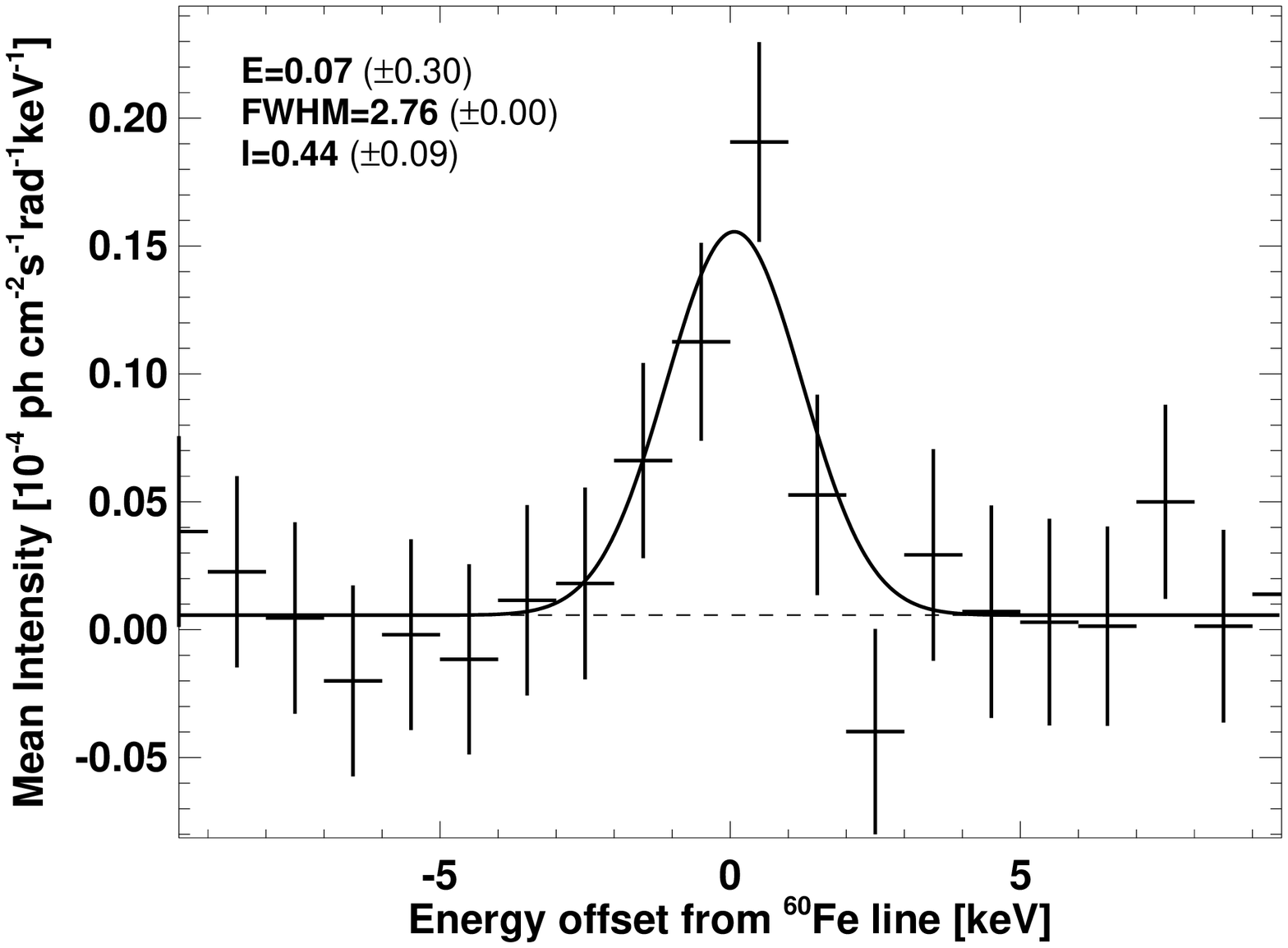}
\qquad
\includegraphics[width=0.47\textwidth]{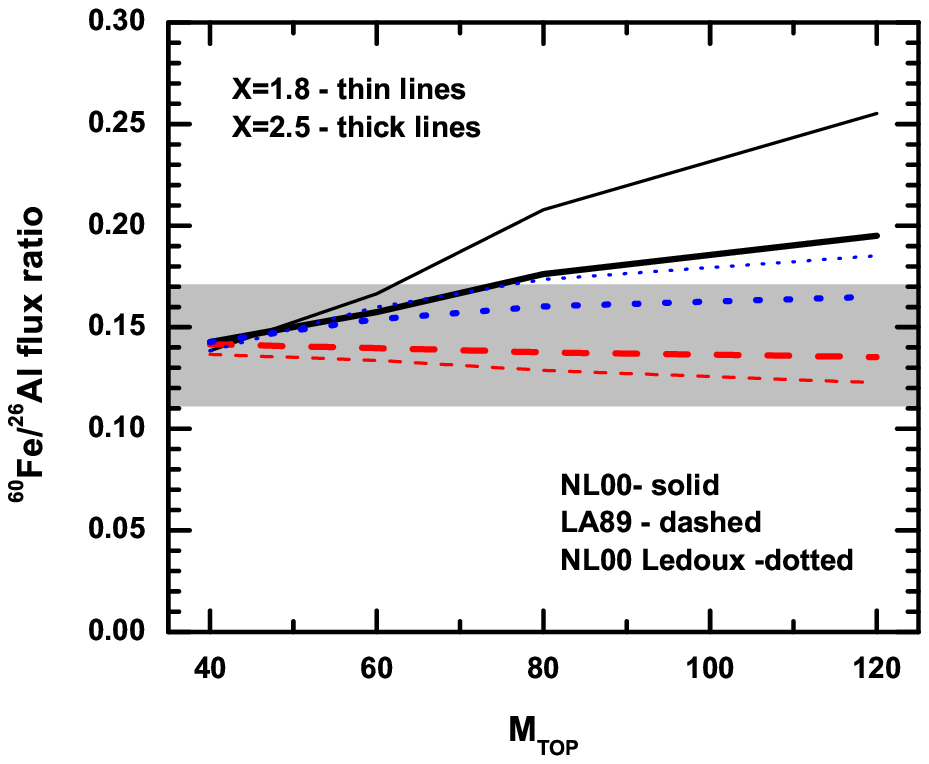}
\caption{Left: \Fh \ line profile from SPI/INTEGRAL observations of the inner Galaxy 
(from Wang et al. 2007). Right: 
theoretical estimates of the  \Fh/\Al \ ratio as a function of the upper mass limit of the stellar initial mass function (curves, based on various assumptions abouth the physics of massive stars) compared to observations (shaded aerea), from Chieffi and Limongi (2006).  }
\label{fig:Fe60}
\end{figure}

\section{Positron annihilation in the Galaxy}

The first $\gamma$-ray line ever detected outside the solar system was the
511 keV line of electron-positron annihilation (Johnson et al. 1972). 
Observations by various instruments in the 90's
established that the line is not variable (at least in a $\sim$10 year period),
that its spatial distribution is apparently dominated by a bulge-like
component and that the overall spectrum suggests a large positronium
fraction of 0.93 (see Kinzer et al. 2001 and references therein). 
The 511 keV flux detected in the central Galactic sterad
was found to be $\sim$10$^{-3}$ ph/cm$^2$/s, corresponding
to a steady state production rate of 10$^{43}$ \ps.

 Observations  in the 2000s with {\it SPI}/INTEGRAL confirmed the abnormally high
 bulge/disk ratio of the 511 keV emission (larger than in any other wavelength, Kn\" odlseder
 et al. 2005) and
 the emission from a disk, albeit with a poorly constrained morphology.
 It is not yet clear whether the disk
is asymmetric (as found in Weidenspontner et al. 2008a) or whether the
bulge centroid is slightly off with respect to the Galactic center (Bouchet et al. 2010)
{\footnote{See also talks by Bouchet, Roques and Skinner in this workshop.}}.

According to the  imaging analysis of {\it SPI} data  (Weidenspointner et al. 2008a)
 the total Galactic e$^+$ annihilation
rate is at least $\dot{N}_{e^+} \sim$2 10$^{43}$ s$^{-1}$, with a
luminosity bulge/disk ratio B/D=1.4. This  model  is further
refined by considering  a narrow ($FWHM=3\degree$)
and a broad ($FWHM=11\degree$) bulge, the former contributing to
$\sim$35\% of the total bulge emission. However,  the data analysis also allows for other 
morphologies,
involving extended regions of low surface brightness but high total
emissivity, e.g. a "halo" of total $\dot{N}_{e^+} \sim$ 3 10$^{43}$
s$^{-1}$ and a thin disk of $\dot{N}_{e^+}\sim$5 10$^{42}$ s$^{-1}$,
leading to a high B/D$\sim$6 (Weidenspointner et al. 2008b).

Information on the origin of those positrons is also obtained via the spectral analysis of the
511 keV emission (Guessoum et al. 2005, Jean et al. 2006, Churazov et al. 2010).
The observed flux at $\sim$MeV energies from the inner Galaxy constrains the initial
energy of the positrons to less than a few MeV (otherwise the emission from in-flight annihilation
would exceed the observed flux, Beacom and Yuksel 2006). Moreover, the spectral analysis provides important information
on the physical properties of the e$^+$ annihilation sites. The large positronium fraction
$f_{P_S}\sim$94-97 \% implies that positrons annihilate mostly at low energies, since direct
annihilation cross-sections are important only at high energies. The overall spectral
shape suggests that annihilation occurs mostly in warm (T$\sim$8 000 K) media, at about equal
amounts in neutral and ionized phases 
but it cannot be  excluded that less than 23\% of annihilation occurs in the cold  
neutral medium (T $\sim$ 80 K); annihilation in the neutral  
media may account for the presence of a broad 511 keV line component  
(FWHM $\sim$5 keV) and the annihilation in the warm ionized medium for  
the narrow one (FWHM $\sim$1 keV).

Among the various astrophysical sources of positrons proposed so
far, the only one known with certainty to release e$^+$ in the ISM
is $\beta^+$ radioactivity of $^{26}$Al; the observed intensity of
its characteristic 1.8 MeV emission in the Galaxy corresponds to
$\sim$3-4 10$^{42}$ e$^+$ s$^{-1}$. A similar amount
is expected from the decay of $^{44}$Ti, on the grounds of
nucleosynthesis arguments. Both radionuclides are
produced mostly in massive stars and their positrons should be
released along the Galactic plane, as traced by the 1.8 MeV
emission; they could thus account for the observed disk 511 keV
emission.

Radioactivity of $^{56}$Co from SNIa was traditionally considered
to be the major e$^+$ producer in the Galaxy. Both the typical
$^{56}$Ni yield of a SNIa and the Galactic SNIa rate are rather
well constrained, resulting in 5 10$^{44}$ e$^+$ s$^{-1}$ produced
{\it inside} SNIa. If only f$_{esc}\sim$4\% of them escape the
supernova to annihilate in the ISM, the observed total e$^+$
annihilation rate can be readily explained. However, observations
of two SNIa, interpreted in the framework of 1-D (stratified)
models, suggest that the positron escape fraction is negligible
{\it at late times}. On the other hand, both observations of early
spectra and 3-D models of SNIa suggest that a sizeable fraction of
$^{56}$Ni is found at high velocity (close to the surface), making
- perhaps - easier the escape of $^{56}$Co positrons. 
In our opinion, SNIa remain a serious candidate, with a
potential Galactic yield of  2 10$^{43}$ e$^+$ s$^{-1}$. But the
expected spatial distribution of SNIa in the Galaxy  corresponds
to a much smaller B/D ratio than that of the observed 511 keV
profile (see Prantzos et al. 2010 for a thorough discussion of SNIa issues
in the context of Galactic positrons).

\begin{figure}
\includegraphics[width=0.49\textwidth]{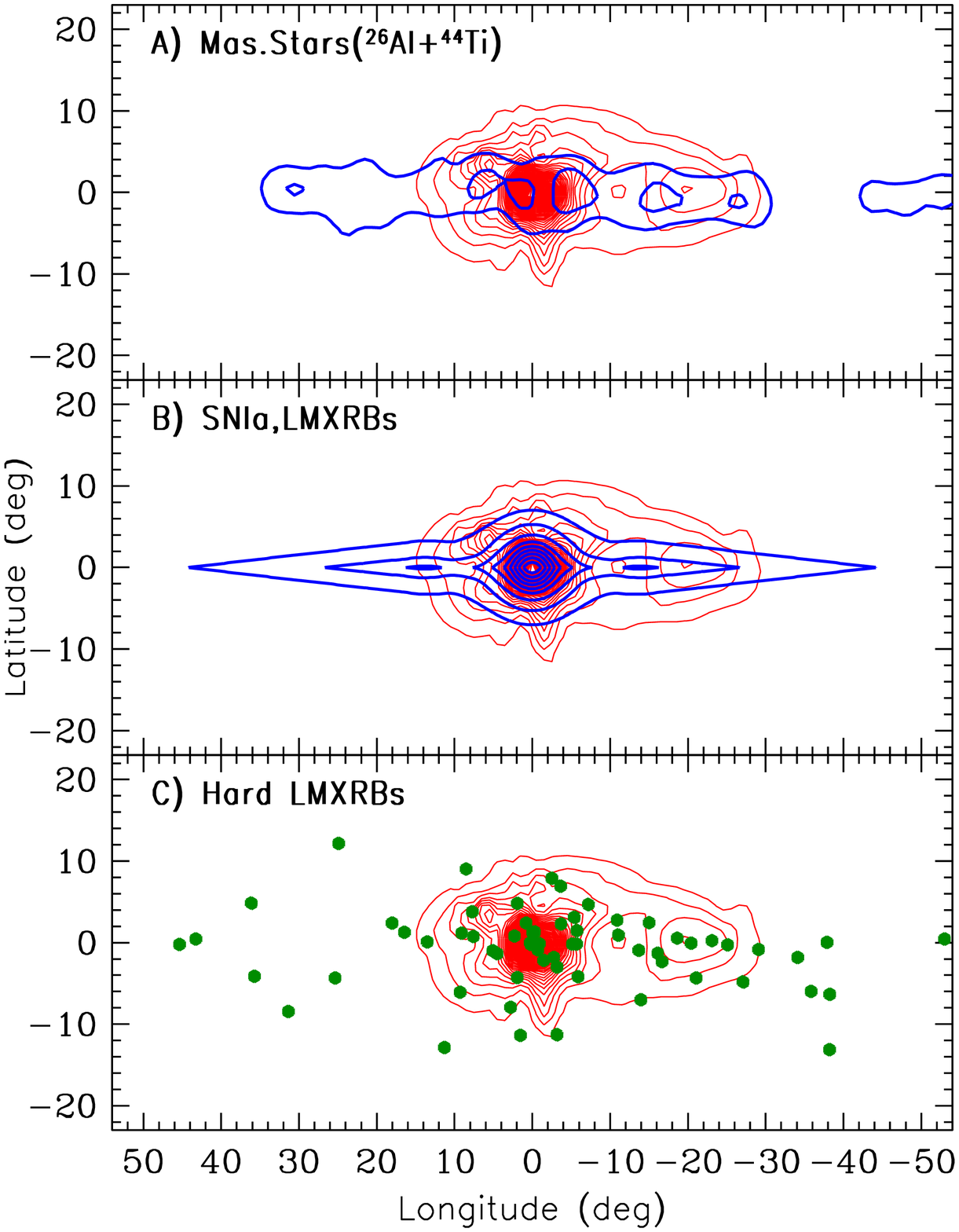}
\qquad
\includegraphics[width=0.49\textwidth]{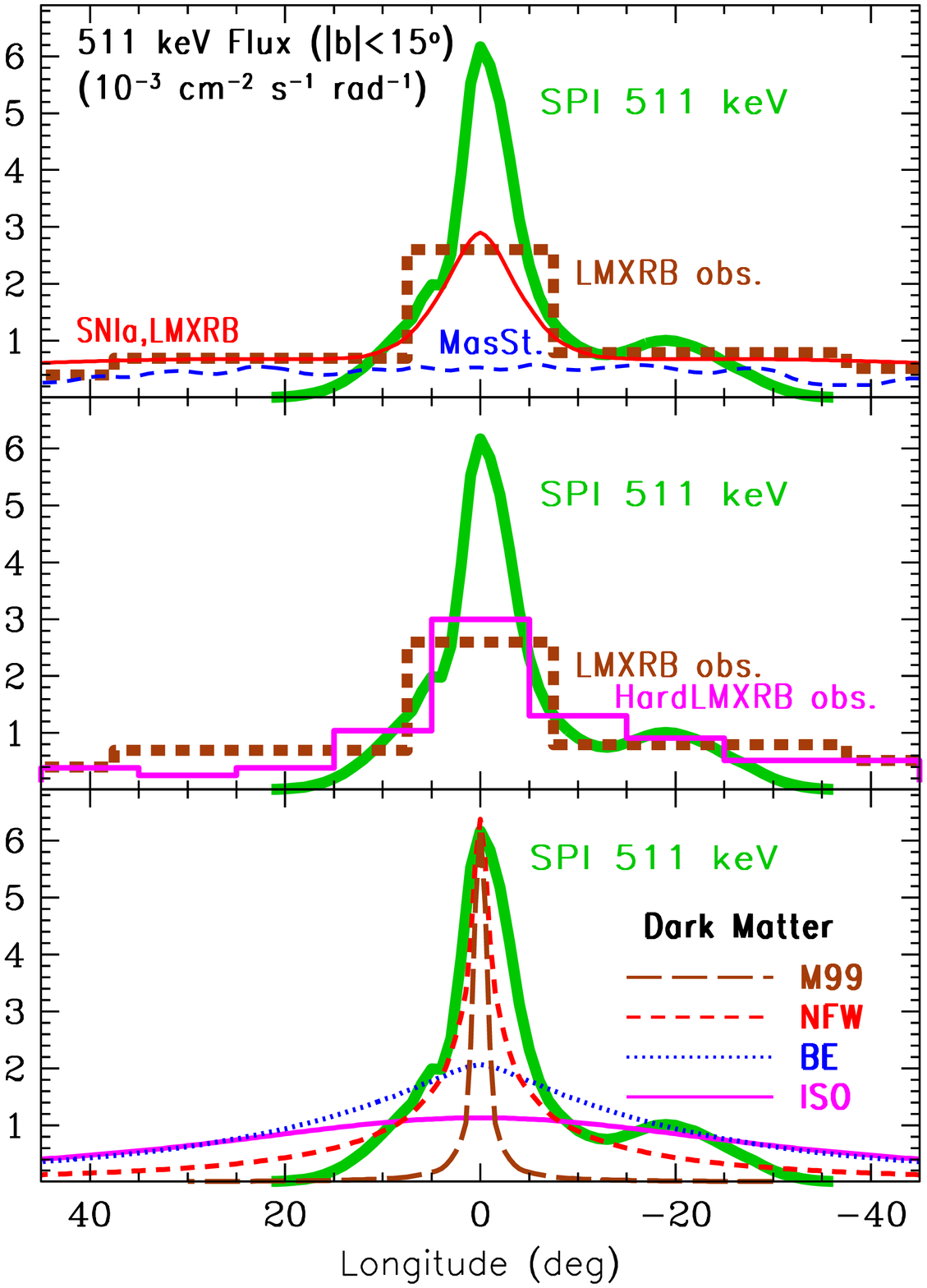} 
\caption{\label{fig:epsart} {\bf Left:} Maps of the Galactic 511 keV emission (flux in cm$^{-2}$ s$^{-1}$ sterad$^{-1}$), 
as observed from SPI (in all panels, {\it thin isocontours}
from Weidenspointneret al. 2008a) and from observationally based or theoretical estimates. A) Observed 
\Al \ (and, presumaby,  $^{44}$Ti) map (from Pluschke et al. 2001 ) ; 
B) Accreting binary systems (SNIa and, presumably, LMXRBs, see text); C) 
Observed Hard LMXRBs. The robustly expected e$^+$ annihilation from radioactivity in the disk 
(upper panel) is not yet fully  seen by SPI.
{\bf Right:} Intensity of 511 keV emission as a function of Galactic longitude.
All fluxes  are integrated for latitudes $|b| <$15$^o$. In all panels, the {\it thick solid curve} corresponds to  SPI observations, i.e. the map of left figure. ({\it Note}: We emphasize that SPI maps and fluxes are provided here for illustration purposes only; quantitative comparison of model predictions to data should only be made through convolution with SPI response matrix.). 
The {\it thick dotted histogram} ({\it top} and {\it middle}) is the observed longitude distribution of LMXRBs (from Grimm et al. 2002);  the latter resembles  closely the theoretically estimated  longitude distribution of SNIa ({\it thin solid curve} in the {\it upper panel}), which  has been normalised to a total emissivity of 1.6 10$^{43}$ e$^+$ s$^{-1}$, with Bulge/Disk=0.45 (maximum Bulge/Disk ratio for SNIa). Also, in the upper panel, the {\it lower dashed  curve}  corresponds to the expected contribution of the $^{26}$Al and $^{44}$Ti $\beta^+$-decay from massive stars. 
The {\it thin solid histogram} in the {\it middle panel} is the observed longitude distribution of Hard LMXRBs (from  Bird et al. 2007) and it has the same normalization as the thick histogram. In the {\it bottom} panel, the SPI 511 keV profile is compared to profiles expected from dark matter annihilation). 
Both figures are from the review of Prantzos et al. (2010).
}
\label{Flux_Image}
\end{figure}

Most of  the other astrophysical candidates   can be constrained to be only minor e$^+$ sources, on the basis of either
weak e$^+$ yields (novae, Galactic cosmic rays),  high e$^+$ energy (compact objects, like
pulsars or  magnetars), spatial morphology of sources (hypernovae, gamma ray bursts) or a combination
of those features (e.g. cosmic rays). Only two astrophysical candidates remain as
potentially important contributors: LMXRBs (Prantzos 2004a) 
or the microquasar variant of that class of sources (Guessoum et al. 2006) 
and the supermassive black hole at the Galactic center (e.g. Cheng et al. 2006, Totani 2006, Chernysov et al. 2009 and references therein). It should be stressed that there is no evidence that either of those sources produces positrons and the e$^+$ yields evaluated by various authors are close 
to upper limits rather than typical
values. Furthermore, because of the current low activity of the central MBH (much lower than that  of LMXRBs) it has to be assumed that the source was much more active in the past, thus dropping the
assumption of "steady state" between e$^+$ production and annihilation,  which is likely in all other cases.

Dark matter (DM) has been proposed as an alternative e$^+$ source,
at least for the bulge 511 keV emission; in principle, it could
complement disk emission originating from radioactivity of $^{26}$Al
and $^{44}$Ti or $^{56}$Co. Observations of the MeV continuum from the inner
Galaxy constrain the large phase space of DM properties. The mass of
annihilating or decaying DM particles should be smaller than a few
MeV, otherwise their in-flight annihilation would overproduce the
MeV continuum.  Scalar light DM particles
with fermionic interactions still appear as a possible candidate (e.g. Boehm et al. 2004);
alternatively, the collisional de-excitation of heavy (100 GeV) DM
particles could provide the required positrons, provided the energy
separation between their excited levels is in the MeV range (e.g. Finkbeiner and Weiner 2007).
 On the other hand, the observed spatial profile of the 511
keV emission constrains the production mode of DM positrons, {\it if
it is assumed that they annihilate close to their production
region}: only "cuspy" profiles are allowed in the case of
annihilating or de-exciting DM particles (for which
$\rho_{\gamma}\propto \rho_{DM}^2$), while decaying DM particles
(for which $\rho_{\gamma}\propto \rho_{DM}$) are excluded; 
the problem is that observations of external galaxies
suggest rather flat, not cuspy, DM profiles.

Positrons produced in the hot, tenuous plasma filling the bulge
(either from SNIa, LMXRBs or DM), have to travel long distances
before slowing down and annihilating (Jean et al. 2006, 2009). This is corroborated by the
spectral analysis, which suggests that positrons annihilate in warm
gas: such gas is filling mostly the inner bulge. Positron
propagation appears then unavoidable, undermining the assumption
that the e$^+$ production and annihilation profiles are correlated,
at least in the bulge. A similar situation should hold for positrons
produced away from the plane of the disk (i.e. from SNIa or LMXRBs),
which is also dominated by hot, tenuous gas. The situation is less
clear for positrons produced by massive star radioactivity, in the
plane of the disk and inside spiral arms: although some of them may
fill hot bubbles and cavities created by the SN explosions and
ultimately escape from the disk, another fraction may annihilate in
closeby dense molecular clouds. Propagation of MeV positrons in the
ISM may then hold the key to understanding the 511 keV emission. It
depends on the physical properties of the ISM (density, ionization) 
but also on the properties of turbulence and magnetic field
configuration. Preliminary attempts to evaluate the extent
of positron propagation and their implications for the
Galactic 511 keV emission (Prantzos 2006, Higdon et al. 2009)  are promising in that respect,
but the situation is far from clear at present: the entanglement
between the various uncertainties (concerning e$^+$ sources, e$^+$
propagation and annihilation sites) does not allow any strong
conclusions to be drawn.

More than 30 years after its discovery, the origin of the first extra-solar $\gamma$-ray line remains unknown. Most probably, observations with next generation instruments will be required to unravel
its mystery.

\def\apj{ApJ}
\def\apjs{ApJSup.}
\def\aa{AA}

{}

\end{document}